\documentclass[12pt]{article}
\usepackage{amsmath}
\usepackage{amsfonts}
\usepackage{amssymb}
\usepackage{amstext}
\usepackage{graphicx}
\usepackage{enumerate}
\usepackage{epsfig}%
\title{\centerline
\bf Length contraction in Very Special Relativity}
\bigskip
\author{Biplab Dutta$^\dagger$, Kaushik Bhattacharya$^\ddagger$
\thanks{email:\,\,$^\dagger$bdutta@iitk.ac.in, $^\ddagger$kaushikb@iitk.ac.in}
\\
\normalsize
Department of Physics, Indian Institute of Technology, Kanpur\\
\normalsize
Kanpur 208016, India
}
\begin{document}
\maketitle
\begin{abstract}
 Glashow and Cohen  claim that many results of special
 theory of relativity (SR) like time dilation, relativistic velocity
 addition, etc, can be explained by using certain proper subgroups, of
 the Lorentz group, which collectively form the main body of Very
 special relativity (VSR). They did not mention about length
 contraction in VSR. Length contraction in VSR has not been studied at
 all. In this article we calculate how the length of a moving rod
 contracts in VSR, particularly in the $HOM(2)$ version. The results
 are interesting in the sense that in general the length contraction
 formulas in VSR are different from SR but in many cases the two
 theories predict similar length contraction of moving rods.
\end{abstract}
\section{Introduction}
In the recent past Glashow and Cohen \cite{glashow} proposed the
interesting idea of a very special relativity.  By very special
relativity (VSR) the above mentioned authors meant a theory which is
constituted by subgroups of the Lorentz group, but amazingly, these
subgroup transformations keep the velocity of light invariant in
inertial frames and time-dilation remains the same as in special
relativity (SR). Velocity addition has been studied in VSR theories
\cite{das} and there has been an attempt to utilize the VSR theory as
the theory of space-time transformations of dark matter candidates
\cite{alhu}. In a parallel development some authors have attempted to
incorporate the framework of VSR in non-commutative space-times
\cite{jabbar}.

The specialty of VSR is that it can produce the constancy of light 
velocity and time-dilation with much smaller subgroups of the Lorentz
group.  Going by standard convention where ${\bf K}$ specify the boost
generators and ${\bf J}$ specify the angular momentum generators of
the full Lorentz group, there are four subgroups of the Lorentz group
which exhausts all the candidates of VSR.  One of the four possible
versions of a theory of VSR has just two generators, namely
$T_1=K_x+J_y$ and $T_2=K_y-J_x$. This group is called $T(2)$. If in
addition to the above generators of $T(2)$ one adds another generator
$J_z$ then the resulting group is called $E(2)$. Instead of adding
$J_z$ if one includes $K_z$ as the third generator in addition to the
two generators of $T(2)$ one attains another subgroup of the Lorentz
group which is called $HOM(2)$. Lastly, if some one includes both
$J_z$ and $K_z$ in addition to the two generators of $T(2)$ then one
obtains another subgroup of the Lorentz group which is called the
$SIM(2)$. These above four subgroups of the Lorentz group which admits
of local energy-momentum conservation collectively form the main body
of VSR transformations.

The topic which still remains untouched is related to the topic of
length contraction in VSR. Till now none of the papers on VSR has
clearly stated about the way how lengths of moving rods differ from
their proper length . In this article we discuss about length
transformations in VSR. To do so we use the subgroup $HOM(2)$.  This
subgroup preserves similarity transformations or homotheties.  It is
seen that the length transformation formulas in VSR are dramatically
different from the one we have in SR. In VSR we observe length
contraction but this contraction is not equivalent to the one found in
SR. More over length-contraction in VSR is direction dependent. If a
rod is placed along the $z$-direction of the fixed frame $S$ and the
moving frame, $S'$, moves with an uniform velocity along the $z$ -
$z'$ axes then the length transformation formula in VSR is exactly the
same as in SR.  But if the rod is placed along the $x$-axis ($y$-axis)
in the $S$ frame and the $S'$ frame moves with an uniform velocity
along $x$ - $x'$ axes ( $y$ - $y'$ axes) then the length
transformation relation is not the same as found in SR. More over for
very high velocities there is no length-contraction along the $x$ or
$y$ axes motion.

The other important find in this article is that the phenomenon of
length contraction is not symmetrical in the frames $S$ and $S'$. By
symmetrical we mean that if the rod is kept at rest in the $S'$ frame,
which is moving with respect to frame $S$ with velocity ${\bf u}$, and
the observer is in the $S$ frame then the length contraction results
does not in general match with the case where the rod is at rest in
the $S$ frame and the observer is in $S'$ frame. This phenomenon
arises because the VSR transformation which links the coordinates of
the primed frame to the unprimed frame is not the same as the inverse
transformation with the sign of the velocity changed. The results
presented in the article can be experimentally tested in heavy ion
collisions and future experiments in LHC. The experiments can
conclusively state whether VSR can actually replace SR in describing
the subtleties of nature.
 
The material in this article is presented in the following format. The
next section explains the VSR transformation, particularly the
$HOM(2)$ version.  The notations and conventions are introduced and
using them the expressions of the $HOM(2)$ transformation and its
inverse transformations are deduced. Section \ref{timedil} deals with
the particular question of length contraction in VSR. The ultimate
section \ref{disc} presents the conclusion with a brief discussion of
the results derived in this article. For the sake of completeness we
attach two appendices as appendix \ref{app1} and appendix \ref{app2}
where the $HOM(2)$ transformation matrix and its inverse are derived
explicitly.
\section{Space-time transformations in the $HOM(2)$ group}
\label{very}
The \textit{HOM}(2) subgroup of the Lorentz group consists of 3
generators $T_1 = K_x + J_y, T_2 = K_y - J_x$ and $K_z$ where $K_i$'s
and $J_i$'s are the generators of Lorentz boosts and 3-space rotations
respectively. The \textit{HOM}(2) generators are $T_1$, $T_2$ and
$K_z$, satisfying the following commutation relations \cite{das},
\begin{equation}
[T_1, T_2] = 0 ,\hspace{10pt} [T_1, K_z] = i T_1,\hspace{10pt} 
[T_2, K_z] = i T_2\,.
\label{bcomm}
\end{equation}
In VSR if one transforms from the rest frame of a particle to a moving
frame, moving with a velocity ${\bf u}$ with respect to the other
frame, the 4-velocity of the particle gets transformed. If the
4-velocity of the particle in the rest frame is $u_0$ and its
4-velocity in the moving frame be $u$ then the 4-vectors must be like
\begin{eqnarray}
u_0 =
\left(
\begin{array}{c}
1\\
0\\
0\\
0
\end{array}
\right)\,,\,\,\,\,
u =
\left(
\begin{array}{c}
\gamma\\
-\gamma u_x\\
-\gamma u_y\\
-\gamma u_z
\end{array}
\right)\,,
\label{conv}
\end{eqnarray}
where $\gamma_u={1}/{\sqrt{1-{\bf u}^2}}$ and ${\bf u}^2=u_x^2 + u_y^2
+ u_z^2$. The $HOM(2)$ transformation acts as:
\begin{equation}\label{1}
L(u) u_0 = u\,,
\end{equation}
where the VSR transformation matrix,
$L(u)$, is given by the following equation
\begin{equation}\label{lu}
L(u) = e^{i\alpha T_1} e^{i\beta T_2} e^{i\phi K_z}.
\end{equation}
The negative sign of the 3-vector part of the 4-vector given in
Eq.~(\ref{conv}) is chosen such that the sign matches to the
corresponding 3-vector in SR. This sign convention is different from
the sign convention used in Ref.~\cite{das, glashow}. The
appropriateness of our sign convention will be discussed once we write
$L(u)$ in the matrix form. The parameters $\alpha$, $\beta$ and $\phi$
are the parameters specifying the transformation and they are given as
\begin{eqnarray}
\label{vi} \alpha &=& -\frac{u_x}{1+u_z}\,,\\
\label{vii} \beta &=& -\frac{u_y}{1+u_z}\,,\\
\label{viii} \phi &=& -\ln (\gamma _u (1+u_z))\,,
\end{eqnarray}
as specified in Ref.~\cite{das, glashow}. The parameters specified
above can be found out by using the form of the 4-vectors in
Eq.~(\ref{conv}) and the VSR transformation equation in
Eq.~(\ref{1}). An explicit derivation of the above parameters is given
in appendix \ref{app1}.

The form of the matrices corresponding to the three transformations
$e^{i\alpha T_1}$, $e^{i\beta T_2}$ and $e^{i\phi K_z}$ can be
calculated by using the standard representations of ${\bf J}$ and
${\bf K}$. The following matrices encapsulate all the properties of
the VSR transformations:
\begin{eqnarray}
e^{i\alpha T_1} &=& \left(
\begin{array}{rrrr}
1+\frac{\alpha^2}{2 !} & \alpha & 0 & -\frac{\alpha^2}{2 !} \\
\alpha & 1 & 0 & -\alpha \\
0 & 0 & 1 & 0 \\
\frac{\alpha^2}{2 !} & \alpha & 0 & 1-\frac{\alpha^2}{2 !}
\end{array}
\right)\,,
\nonumber\\
e^{i\beta T_2} &=& \left(
\begin{array}{rrrr}
1+\frac{\beta^2}{2 !} & 0 & \beta & -\frac{\beta^2}{2 !} \\
0 & 1 & 0 & 0 \\
\beta & 0 & 1 & -\beta \\
\frac{\beta^2}{2 !} & 0 & \beta & 1-\frac{\beta^2}{2 !}
\end{array}
\right)\,,
\nonumber\\
\label{eiphi}
e^{i\phi K_z} &=& \left(
\begin{array}{rrrr}
\cosh\phi & 0 & 0 & \sinh\phi \\
0 & 1 & 0 & 0 \\
0 & 0 & 1 & 0 \\
\sinh\phi & 0 & 0 & \cosh\phi
\end{array}
\right)\,.
\nonumber
\end{eqnarray}
From the form of the three transformations as listed above a general
$HOM(2)$ transformation, $L(u)$ as given in Eq.~(\ref{lu}), written in
terms of the velocity components $u_x$, $u_y$ and $u_z$, can be written as,
\begin{equation}
L(u)=\left(
\begin{array}{cccc}
 \gamma _u & -\frac{u_x}{1+u_z} & -\frac{u_y}{1+u_z} & 
-\frac{\gamma _u\left(u_z + {\bf u}^2\right)}{\left(1+u_z\right)} \\
 -\gamma _uu_x & 1 & 0 & \gamma _uu_x \\
 -\gamma _uu_y & 0 & 1 & \gamma _uu_y \\
 -\gamma _uu_z & -\frac{u_x}{1+u_z} & -\frac{u_y}{1+u_z} & 
\frac{\gamma _u\left(1- {\bf u}^2+u_z+u_z{}^2\right)}{\left(1+u_z\right)}
\end{array}
\right)\,.
\label{mglu}
\end{equation}
In the above expression of $L(u)$ if we put $u_x=0$ and $u_y=0$ then
we get a resultant $L(u_z)$ which is equivalent to the Lorentz
transformation matrix in SR. The signs of the resulting SR
transformation matrix matches with the sign of our $L(u_z)$. This
matching of $L(u_z)$ with the corresponding SR Lorentz transformation
matrix dictates the sign convention of the 3-vector of u in
Eq.~(\ref{conv}). The convention of the SR Lorentz transformation
followed in this article matches with the convention of Landau and
Lifshitz as they explained it in Ref.~\cite{landau}.

In a similar way one can also calculate the inverse of the $HOM(2)$
transformation, $L^{-1}(u)$, where the inverse transformation is defined as 
\begin{equation}
L^{-1}(u) u = u_0\,.
\end{equation}
In the present case,
\begin{equation}
L^{-1}(u) =  e^{-i\phi K_z} e^{-i\beta T_2} e^{-i\alpha T_1}\,.
\label{minuslu}
\end{equation}
The individual transformation matrices now are given as:
\begin{eqnarray}
e^{-i\phi K_z} &=& \left(
\begin{array}{rrrr}
\cosh\phi & 0 & 0 & -\sinh\phi \\
0 & 1 & 0 & 0 \\
0 & 0 & 1 & 0 \\
-\sinh\phi & 0 & 0 & \cosh\phi
\end{array}
\right)\,,
\nonumber\\
e^{-i\beta T_2} &=& \left(
\begin{array}{rrrr}
1+\frac{\beta^2}{2 !} & 0 & -\beta & -\frac{\beta^2}{2 !} \\
0 & 1 & 0 & 0 \\
-\beta & 0 & 1 & \beta \\
\frac{\beta^2}{2 !} & 0 & -\beta & 1-\frac{\beta^2}{2 !}
\end{array}
\right)\,,
\nonumber\\
e^{-i\alpha T_1} &=& \left(
\begin{array}{rrrr}
1+\frac{\alpha^2}{2 !} & -\alpha & 0 & -\frac{\alpha^2}{2 !} \\
-\alpha & 1 & 0 & \alpha \\
0 & 0 & 1 & 0 \\
\frac{\alpha^2}{2 !} & -\alpha & 0 & 1-\frac{\alpha^2}{2 !}
\end{array}
\right)\,,
\nonumber
\end{eqnarray}
where the parameters $\alpha, \beta, \phi$ are given by (\ref{vi}),
(\ref{vii}), and (\ref{viii}). The inverse transformation matrix of
the ${HOM}(2)$ group is given as,
\begin{equation}
\label{luinv}
L^{-1}(u) = \left(
\begin{array}{cccc}
 \gamma _u & \gamma _uu_x & \gamma _uu_y & \gamma _uu_z \\
 \frac{u_x}{1+u_z} & 1 & 0 & -\frac{u_x}{1+u_z} \\
 \frac{u_y}{1+u_z} & 0 & 1 & -\frac{u_y}{1+u_z} \\
 \frac{\gamma _u\left(u_z + {\bf u}^2\right)}{\left(1+u_z\right)} 
& \gamma _uu_x & \gamma _uu_y & \frac{\gamma _u
\left(1- {\bf u}^2+u_z+u_z{}^2\right)}{\left(1+u_z\right)}
\end{array}
\right)\,.
\end{equation}
It can be shown that with these forms of $L(u)$ and $L^{-1}(u)$ one
obtains $L(u)L^{-1}(u)=L^{-1}(u)L(u)=1$. From the expressions in
Eq.~(\ref{mglu}) and Eq.~(\ref{luinv}) it is clear that the inverse
transformation in VSR is not obtained by altering the signs of the
velocity components in the $L(u)$ matrix. This property of the VSR
transformations differ from the corresponding property of SR
transformations.  Putting $u_x=0$ and $u_y=0$ in the expression of
$L^{-1}(u)$ we get a resultant $L(u_z)$ which is equivalent to the
corresponding inverse Lorentz transformation matrix in SR.

Let us consider two inertial frames $S'$ and $S$ which coincide with
each other at $t=t'=0$. Suppose the $S'$ frame is moving with velocity
${\bf u}$ with respect to $S$ frame.  The coordinates of the two frames are
related by
\begin{equation}
\mathbf{x} = L^{-1}(u)\, \mathbf{x'}\,,
\end{equation}
where $\mathbf{x}=(t, x, y, z)$ and $\mathbf{x'} = (t', x', y', z')$.
Using Eq.~(\ref{luinv}) the coordinate transformation equations
can be explicitly written as,
\begin{eqnarray}
\label{time}t &=& \gamma _u t'+ \gamma _uu_x x'+  \gamma _uu_y y'+ 
\gamma _uu_z z'\\
\label{xspace}x &=& \frac{u_x}{1+u_z} t'+ x'- \frac{u_x}{1+u_z} z'\\
\label{yspace}y &=& \frac{u_y}{1+u_z} t'+ y' - \frac{u_y}{1+u_z} z'\\
\label{zspace}z &=& \frac{\gamma _u\left(u_z+ {\bf u}^2\right)}
{\left(1+u_z\right)} t' + \gamma _uu_x x' + \gamma _uu_y y' + 
\frac{\gamma _u\left(1- {\bf u}^2+u_z+u_z^2\right)}{\left(1+u_z\right)} z'
\end{eqnarray}
From the above equations one can explicitly verify that $ds^2={ds'}^2$,
where the invariant line-element squared is ${ds}^2 = - {dt}^2 +
{dx}^2 + {dy}^2 + {dz}^2$.  As the square of the line-element remains
invariant under $HOM(2)$ transformations one can derive a time
dilation formula in this case. The result matches exactly with that of SR.
\section{Length of a moving rod in VSR}
\label{timedil}
In this section we will discuss the length contraction formulas in
VSR.  As VSR does have a preferred direction, which is along the
$z$-axis of the $S$ frame, one cannot arbitrarily rotate the
coordinate systems to suit ones need as in SR. In this case one has to
rely more on mathematical description of the physical problem and the
concept of isotropy has to be kept aside. The most general treatment
of the length contraction formula requires the moving rod to be
arbitrarily placed in the moving frame which can have any arbitrary
velocity (although the magnitude of velocity must be smaller than
one).  The general setting of the length contraction problem is too
complicated and cumbersome in VSR as because the transformation
equations are themselves complicated as compared to SR. But a
meaningful approach and some interesting results can be obtained by
some specific cases and in this section we will try to elucidate
these points explicitly.
\subsection{The rod is at rest in the $S$ frame}
In this section we discuss the issue about length transformations in
VSR.  We will focus our attention particularly to the ${HOM}(2)$
transformations. For the first case we suppose that a rod is at rest
along the $x$-axis in the $S$ frame. The length of the rod is $\Delta
x=x_2 - x_1 \equiv l_0 $. An observer in the $S'$ frame, which is
moving with a velocity ${\bf u}$ with respect to the $S$ frame, can
measure the length of the rod in his/her frame. For the measurement of the 
length of the rod in motion one has to know the coordinates 
of the two ends of the rod (($x'_1, y'_1, z'_1$) and ($x'_2, y'_2, z'_2$))
simultaneously (at $t'$). From the form of the coordinate transformations 
given in the last section we can write that
\begin{eqnarray}
x_1 &=& \frac{u_x}{1+u_z} t'+ x'_1 - \frac{u_x}{1+u_z} z'_1\,,
\nonumber\\
y_1 &=& \frac{u_y}{1+u_z} t' + y'_1 - \frac{u_y}{1+u_z} z'_1\,,
\nonumber\\
z_1 &=& \frac{\gamma _u\left(u_z+ {\bf u}^2\right)}
{\left(1+u_z\right)} t' + \gamma _uu_x x'_1 + \gamma _uu_y y'_1 + 
\frac{\gamma _u\left(1- {\bf u}^2+u_z+u_z^2\right)}{\left(1+u_z\right)} z'_1\,,
\nonumber
\end{eqnarray}
which connects the coordinates of one end of the rod in $S$ frame to
its corresponding coordinates in the $S'$ frame.  A similar set of
relations can be written for $x_2$, $y_1$ and $z_1$ (coordinates of
the other end of the rod in $S$ frame) as
\begin{eqnarray}
x_2 &=& \frac{u_x}{1+u_z} t'+ x'_2 - \frac{u_x}{1+u_z} z'_2\,,
\nonumber\\
y_1 &=& \frac{u_y}{1+u_z} t' + y'_2 - \frac{u_y}{1+u_z} z'_2\,,
\nonumber\\
z_1 &=& \frac{\gamma _u\left(u_z+ {\bf u}^2\right)}
{\left(1+u_z\right)} t' + \gamma _uu_x x'_2 + \gamma _uu_y y'_2 + 
\frac{\gamma _u\left(1- {\bf u}^2+u_z+u_z^2\right)}{\left(1+u_z\right)} z'_2\,.
\nonumber
\end{eqnarray}
It is interesting to note that although $y$ and $z$ coordinates remain
the same for the two ends of the rod in the $S$ frame, in the $S'$
frame it does not remain so. Subtracting the first triplet of
equations from the second triplet we have
\begin{eqnarray}
l_0 &=& \Delta x' - \frac{u_x}{1+u_z}\Delta z'\,,
\label{trans1}\\
0 &=& \Delta y' - \frac{u_y}{1+u_z}\Delta z'\,,
\label{trans2}\\
0 &=& \gamma _uu_x \Delta x'  + \gamma _uu_y \Delta y' + 
\frac{\gamma _u\left(1- {\bf u}^2+u_z+u_z^2\right)}{\left(1+u_z\right)} 
\Delta z' \,,
\label{trans3}
\end{eqnarray}
where $\Delta x'\equiv x'_2 - x'_1$, $\Delta y'\equiv y'_2 - y'_1 $
and $\Delta z'\equiv z'_2 - z'_1$. Using the first two equations, in the
above set of equations, one can deduce from Eq.~(\ref{trans3}) that
\begin{eqnarray}
\Delta z' = -u_x l_0\,.
\end{eqnarray}
Using Eq.~(\ref{trans1}), Eq.~(\ref{trans2}) and the above equation
one can show that
\begin{eqnarray}
 l = l_0\sqrt{\left(1-\frac{u_x^2}{1+u_z}\right)^2 + \frac{u_x^2u_y^2}
{\left(1+u_z\right)^2} + u_x^2}\,,
\label{xlc}
\end{eqnarray}
where 
\begin{eqnarray}
l\equiv \sqrt{(\Delta x')^2 + (\Delta y')^2 + (\Delta z')^2}\,,
\label{lexp}
\end{eqnarray}
is the length of the rod in the $S'$ frame. It may happen that the
rod, at rest, is placed along the $x$ axis of the $S$ frame while some
of the velocity components of ${\bf u}$ are zero. In this case
Eq.~(\ref{xlc}) gets simplified. If $u_y\ne 0$ and $u_z=u_x=0$, or
$u_z\ne 0$ and $u_x=u_y=0$, in both cases the length of the rod
remains the same. On the other hand if $u_x\ne 0$ and $u_y=u_z=0$ we
will have length contraction and Eq.~(\ref{xlc}) becomes
\begin{eqnarray}
 l = l_0\sqrt{(1-u_x^2)^2 + u_x^2}\,,
\label{xlcs}
\end{eqnarray}
which shows that in general there will be a length contraction but the
amount of contraction depends on $u_x$. For $u_x \to 1$ length
contraction disappears.

If the rod at rest in the $S$ frame was kept along the $y$ axis with
coordinates $(x_1, y_1, z_1)$ and $(x_1, y_2, z_1)$ then analyzing in
a similar way one can write
\begin{eqnarray}
 l = l_0\sqrt{\left(1-\frac{u_y^2}{1+u_z}\right)^2 + \frac{u_y^2u_x^2}
{\left(1+u_z\right)^2} + u_y^2}\,,
\label{ylc}
\end{eqnarray}
where now $l_0=y_2 - y_1$ and $l$ is as given in Eq.~(\ref{lexp}).  If
$u_x\ne 0$ and $u_y=u_z=0$, or $u_z\ne 0$ and $u_x=u_y=0$, in both
cases the length of the rod remains the same. On the other hand if
$u_y\ne 0$ and $u_z=u_x=0$ we will have a length contraction formula
equivalent to the one in Eq.~(\ref{xlcs}) where $u_x$ is replaced by
$u_y$.  The $HOM(2)$ transformations have a preferred axis which is
along the $z$-axis and consequently the length transformation formulas
along these two directions are similar. 

\begin{figure}[h!]
\centering
\includegraphics[width=13cm,height=8cm]{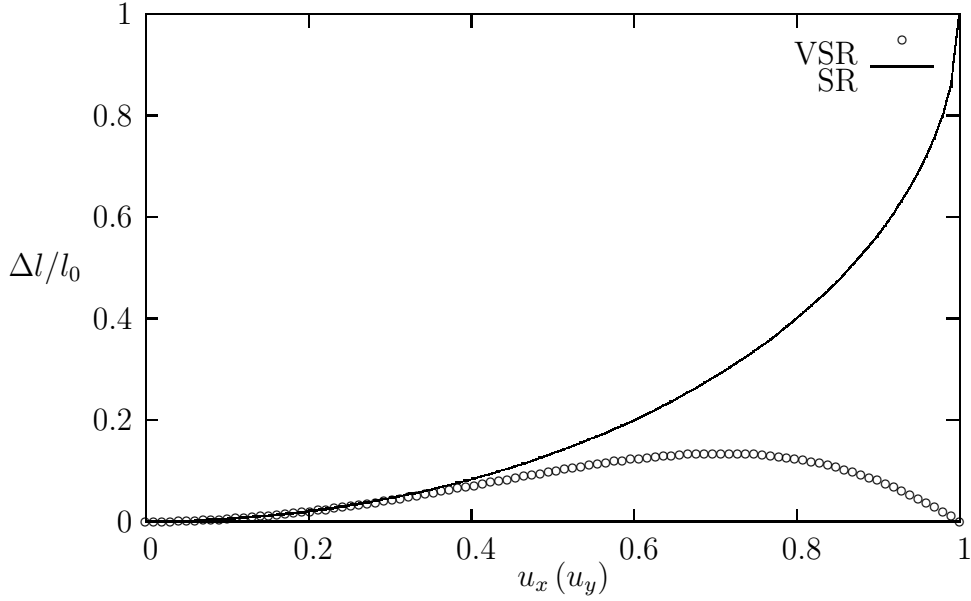}
\caption[]{The plot of the $\Delta l/l_0$, where $\Delta l\equiv l_0 -
  l$, for the case of VSR and SR when the rod is at rest in $S$
  frame. In the abscissa we have either $u_x$ (when $u_y=u_z=0$) or
  $u_y$ (when $u_x=u_z=0$).}
\label{vsrp}
\end{figure}

Lastly we come to the case where the rod at rest in the $S$ frame is
along the $z$ axis with coordinates of its ends given by $(x_1, y_1,
z_1)$ and $(x_1, y_2, z_2)$. In this case if the length of the rod at
rest is given by $l_0=z_2 - z_1$ then, using Eq.(\ref{trans1}),
Eq.~(\ref{trans2}) and Eq.~(\ref{trans3}), it can be shown that
\begin{eqnarray}
 l = l_0\sqrt{\frac{u_x^2(1-{\bf u}^2)}{(1+u_z)^2} + 
\frac{u_y^2(1-{\bf u}^2)}{(1+u_z)^2}+ (1-{\bf u}^2)}\,,
\label{zlc}
\end{eqnarray}
where $l$ is as given in Eq.~(\ref{lexp}). It is immediately observed
that the length transformation formula for the third case is different
from the previous two cases. This has to do with the special status of the 
$z$-axis in $HOM(2)$ and in general in VSR. If
$u_x\ne 0$ and $u_y=u_z=0$, or $u_y\ne 0$ and $u_z=u_x=0$, in both
cases the length of the rod remains the same. On the other hand if
$u_z\ne 0$ and $u_x=u_y=0$ we will have a length contraction formula
equivalent to the one in SR as
\begin{eqnarray}
l=l_0 \sqrt{1-u_z^2}\,. 
\label{srlc}
\end{eqnarray}
This formula, corresponding to relative motion along $z$-axis, again
shows the special status of the preferred axis. If fractional length
contraction is defined by $\Delta l/l_0$ where $\Delta l\equiv l_0 -
l$, for the case of VSR and SR then the contents of Eq.~(\ref{xlcs})
and Eq.~(\ref{srlc}) can be plotted to show the difference of VSR and
SR. Such a plot is given in Fig.~\ref{vsrp}.

It can be checked, using the fact that $|{\bf u}|^2 < 1$, that the
expressions in Eq.~(\ref{xlc}), Eq.~(\ref{ylc}) and Eq.~(\ref{zlc})
gives length contractions. Although VSR transformations constitute
only a small subgroup of the total Lorentz group but here the length
of a moving rod never expands. 
\subsection{The rod is at rest in the $S'$ frame}
In this case we suppose that a rod is at rest along the $x'$-axis in
the $S'$ frame. The length of the rod is $\Delta x'=x'_2 - x'_1 \equiv
l_0 $. An observer in the $S$ frame, which is moving with a velocity
$-{\bf u}$ with respect to the $S$ frame, can measure the length of
the rod in his/her frame. For the measurement of the length of the rod
in motion one has to know the coordinates of the two ends of
the rod (($x_1, y_1, z_1$) and ($x_2, y_2, z_2$)) simultaneously (at
$t$). From the form of $L(u)$, in Eq.~(\ref{mglu}), we can write
\begin{eqnarray}
x'_1 &=& -\gamma_u u_x t + x_1 + \gamma_u u_x z_1\,,
\nonumber\\
y'_1 &=& -\gamma_u u_y t + y_1 + \gamma_u u_y z_1\,, 
\nonumber\\
z'_1 &=& -\gamma_u u_z t - \frac{u_x}{1+u_z} x_1 
- \frac{u_y}{1+u_z} y_1 + 
\frac{\gamma_u\left(1- {\bf u}^2+u_z+u_z^2\right)}{\left(1+u_z\right)} z_1\,,
\nonumber
\end{eqnarray}
which connects the spatial coordinates of one end of the rod in the two frames.
For the the other end we must have
\begin{eqnarray}
x'_2 &=& -\gamma_u u_x t + x_2 + \gamma_u u_x z_2\,,
\nonumber\\
y'_1 &=& -\gamma_u u_y t + y_2 + \gamma_u u_y z_2\,, 
\nonumber\\
z'_1 &=& -\gamma_u u_z t - \frac{u_x}{1+u_z} x_2 
- \frac{u_y}{1+u_z} y_2 + 
\frac{\gamma_u\left(1- {\bf u}^2+u_z+u_z^2\right)}{\left(1+u_z\right)} z_2\,.
\nonumber
\end{eqnarray}
Subtracting the first triplet of
equations from the second triplet we have
\begin{eqnarray}
l_0 &=& \Delta x + \gamma_u u_x \Delta z\,,
\label{ntrans1}\\
0 &=& \Delta y + \gamma_u u_y \Delta z\,,
\label{ntrans2}\\
0 &=& -\frac{u_x}{1+u_z} \Delta x  - \frac{u_y}{1+u_z} \Delta y  + 
\frac{\gamma _u\left(1- {\bf u}^2+u_z+u_z^2\right)}{\left(1+u_z\right)} 
\Delta z \,,
\label{ntrans3}
\end{eqnarray}
where $\Delta x \equiv x_2 - x_1$, $\Delta y \equiv y_2 - y_1 $
and $\Delta z \equiv z_2 - z_1$. From the last set of equations one obtains
\begin{eqnarray}
\Delta z = \frac{l_0 u_x}{\gamma_u (1+u_z)}\,.
\label{delz}
\end{eqnarray}
If the length of the moving rod be 
\begin{eqnarray}
l\equiv \sqrt{(\Delta x)^2 + (\Delta y)^2 + (\Delta z)^2}\,,
\label{newl}
\end{eqnarray}
then in this case we will have
\begin{eqnarray}
 l = l_0\sqrt{\left(1-\frac{u_x^2}{1+u_z}\right)^2 + \frac{u_x^2u_y^2}
{\left(1+u_z\right)^2} + u_x^2\frac{(1-{\bf u}^2)}{(1+u_z)^2}}\,.
\label{nxlc}
\end{eqnarray}
The equation is not equivalent to Eq.~(\ref{xlc}) showing that length
contraction depends upon the frame in VSR. If the rod was kept along
the $y'$ direction in the $S'$ frame, then $l_0= y'_2 - y'_1 $, we
would expect the length contraction formula to be exactly same as that
given above except an interchange in $u_x$ and $u_y$. If $u_x=0$ then
there is no length transformation. If on the other hand $u_x\ne 0$ but
$u_y=u_z=0$ then the above formula becomes
\begin{eqnarray}
 l = l_0\sqrt{1- u_x^2}\,,
\label{nsxlc}
\end{eqnarray}
which matches exactly with the corresponding result from SR. If on the
other hand the rod is placed along the $y'$ axis and $u_y \ne 0$ but
$u_z=u_x=0$ then also we get a length contraction formula exactly
similar to Eq.~(\ref{nsxlc}) except that there $u_x$ is replaced by
$u_y$. In both these cases it is observed that we get the same results
as we get from SR.

If the rod in the $S'$ frame is placed along the $z$ axis it can be
easily found out that in this case
\begin{eqnarray}
l=l_0 \sqrt{u_x^2 + u_y^2 + (1-{\bf u})^2}\,,
\label{zprime}
\end{eqnarray}
where $l$ is as given in Eq.~(\ref{newl}) and $l_0= z'_2 - z'_1 $. If
in this case if $u_z\ne 0$ but $u_x=u_y=0$ we again get back the SR
formula as $l=l_0 \sqrt{1-u_z^2}$. In this case also one can check,
using the fact that $|{\bf u}|^2 < 1$, that Eq.~(\ref{nxlc}) and
Eq.~(\ref{zprime}) gives length contractions.
\section{Discussion and conclusion}
\label{disc}
From the above analysis of length transformations in VSR we see that
in general the VSR results and SR result do not match. But there are
remarkable similarities which may hinder one from discovering the
difference in the results predicted by the two theories. In our
convention there are two frames $S$ and $S'$ which coincides with
each other at $t=t'=0$. As time evolves $S'$ frame moves relative to
$S$ frame with an uniform 3-velocity ${\bf u}$. If the rod is kept
along any axes in the $S'$ frame and one measures its length from $S$
frame then the length transformations do not coincide with the SR
results. But interestingly if the rod is sliding along any common axes
of $S$ and $S'$ frame then we get the exact SR length contraction
results. Consequently if the observer is in $S$ frame and the rod is
oriented along any 3-axes, where the particular axis is along the
direction of the relative velocity ${\bf u}$, one will never discover
whether the theory of relativity is SR or VSR. This difference between
SR and VSR becomes more blurred because as in SR in VSR also the
velocity of light is independent of the reference frame and the
time-dilation formula is exactly the same as in SR.

On the other hand if the rod is at rest in the $S$ frame itself and
the observer is in the $S'$ frame then the length transformation
formulas are different from SR if the orientation of the rod and the
relative velocity is along the $x$ or $y$ axes. But if the rod is
placed along the $z$ axis in the $S$ frame and the frame $S'$ also moves
along the $z$ axis of the $S$ frame with velocity $u_z$ then the
length of the rod measured in the $S'$ frame is contracted in the same
way as one expects from SR.

From the forms of $L(u)$ and $L^{-1}(u)$ as given in Eq.~(\ref{mglu})
and Eq.~(\ref{luinv}) it is observed that $L^{-1}(u)$ is not obtained
from $L(u)$ by putting a minus sign in front of all the velocity
components appearing in the expression of $L(u)$. This is the cause of
different length contraction formulas for two different frames as
shown in the last section. The interesting thing about VSR is that
inspite of these \textit{asymmetric} nature of the transformations,
the square of the line element remains invariant as in SR and
consequently the time-dilation formulas remain exactly the same as in
SR. The length contraction formulas in VSR do depend upon the sign of
$u_z$ which shows that the amount of contraction of length of a moving
rod depends upon its direction of motion along the $z$ or $z'$-axes.

If VSR is really the theory which nature follows, may be in the very
high energy sector or near the electro-weak symmetry breaking
scale, then one may hope to see the effects of VSR length contractions
in the LHC or in future colliders. At present there is no confirmation
of any difference of the length contraction results obtained from
SR. In nearby future heavy ion collision experiments and other related
experiments can really be done to look for any discrepancy of the
length contraction formulas from SR.

In conclusion it must be stated that in this article we have studied
how a moving rod's length changes from its rest length in VSR and
specifically in the $HOM(2)$ version of VSR.  Length contraction is
observed for all the cases but there are some variation in the
transformation equations in contrast to that in SR, although SR
results are reproduced in many special cases of VSR. The other
important conclusion is related to the fact that in general the
phenomenon of length contraction of a rod in VSR do depend upon the
frame from which one observes, a fact which is very difficult to
accept in any relativistic theory.
\appendix\section*{\hfil Appendix \hfil}
\section{$HOM(2)$ transformations}
\label{app1}
The VSR generators are given by $T_1 = K_x + J_y, T_2 = K_y - J_x$ and
$K_z$ where $K_i$'s and $J_i$'s are the generators of Lorentz boosts
and 3-space rotations of the full Lorentz group. In this article we
choose the following form of the generators of ${\bf J}$ and ${\bf K}$
as given in the book by L.~H.~Ryder \cite{ryder}:
\begin{eqnarray}
J_x=-i\left(
\begin{array}{cccc}
0 & 0 & 0 & 0 \\
0 & 0 & 0 & 0 \\
0 & 0 & 0 & 1 \\
0 & 0 & -1 & 0 
\end{array}
\right)\,,\,\,
J_y=-i\left(
\begin{array}{cccc}
0 & 0 & 0 & 0 \\
0 & 0 & 0 & -1 \\
0 & 0 & 0 & 0 \\
0 & 1 & 0 & 0 
\end{array}
\right)\,,\,\,
J_z=-i\left(
\begin{array}{cccc}
0 & 0 & 0 & 0 \\
0 & 0 & 1 & 0 \\
0 & -1 & 0 & 0 \\
0 & 0 & 0 & 0 
\end{array}
\right)\,,
\nonumber
\end{eqnarray}
and 
\begin{eqnarray}
K_x=-i\left(
\begin{array}{cccc}
0 & 1 & 0 & 0 \\
1 & 0 & 0 & 0 \\
0 & 0 & 0 & 0 \\
0 & 0 & 0 & 0 
\end{array}
\right)\,,\,\,
K_y=-i\left(
\begin{array}{cccc}
0 & 0 & 1 & 0 \\
0 & 0 & 0 & 0 \\
1 & 0 & 0 & 0 \\
0 & 0 & 0 & 0 
\end{array}
\right)\,,\,\,
K_z=-i\left(
\begin{array}{cccc}
0 & 0 & 0 & 1 \\
0 & 0 & 0 & 0 \\
0 & 0 & 0 & 0 \\
1 & 0 & 0 & 0 
\end{array}
\right)\,.
\nonumber
\end{eqnarray}
Consequently one must have
\begin{eqnarray}
T_1=-i\left(
\begin{array}{cccc}
0 & 1 & 0 & 0 \\
1 & 0 & 0 & -1 \\
0 & 0 & 0 & 0 \\
0 & 1 & 0 & 0 
\end{array}
\right)\,,\,\,\,
T_2=-i\left(
\begin{array}{cccc}
0 & 0 & 1 & 0 \\
0 & 0 & 0 & 0 \\
1 & 0 & 0 & -1 \\
0 & 0 & 1 & 0 
\end{array}
\right)\,.
\nonumber
\end{eqnarray}
Noting that 
\begin{eqnarray}
T^2_1=-\left(
\begin{array}{cccc}
1 & 0 & 0 & -1 \\
0 & 0 & 0 & 0 \\
0 & 0 & 0 & 0 \\
1 & 0 & 0 & -1 
\end{array}
\right)\,,\,\,\,
T^2_2=-\left(
\begin{array}{cccc}
1 & 0 & 0 & -1 \\
0 & 0 & 0 & 0 \\
0 & 0 & 0 & 0 \\
1 & 0 & 0 & -1 
\end{array}
\right)\,,
\nonumber
\end{eqnarray}
and $T_1^3=T_2^3=0$ we have
\begin{eqnarray}
e^{i\alpha T_1}=\left(
\begin{array}{cccc}
1+\frac{\alpha^2}{2!} & \alpha & 0 & -\frac{\alpha^2}{2!} \\
\alpha & 1 & 0 & -\alpha \\
0 & 0 & 1 & 0 \\
\frac{\alpha^2}{2!} & \alpha & 0 & 1-\frac{\alpha^2}{2!} 
\end{array}
\right)\,,\,\,\,
e^{i\beta T_2}=\left(
\begin{array}{cccc}
1+\frac{\beta^2}{2!} & 0 & \beta & -\frac{\beta^2}{2!} \\
0 & 1 & 0 & 0 \\
\beta & 0 & 1 & -\beta \\
\frac{\beta^2}{2!} & 0 & \beta & 1-\frac{\beta^2}{2!}
\end{array}
\right)\,.
\label{t1t2}
\end{eqnarray}
The square of $K_z$ is given by
\begin{eqnarray}
K_z^2=-\left(
\begin{array}{cccc}
1 & 0 & 0 & 0 \\
0 & 0 & 0 & 0 \\
0 & 0 & 0 & 0 \\
0 & 0 & 0 & 1 
\end{array}
\right)\,,
\nonumber
\end{eqnarray}
and it comes out trivially that $K_z^3=-K_z$. From these facts one can write
\begin{eqnarray}
e^{i\phi K_z }=\left(
\begin{array}{cccc}
\cosh \phi & 0 & 0 & \sinh \phi \\
0 & 1 & 0 & 0 \\
0 & 0 & 1 & 0 \\
\sinh \phi & 0 & 0 & \cosh \phi
\end{array}
\right)\,.
\label{phikz}
\end{eqnarray}
Using the above forms of the matrices we can now calculate 
$L(u) = e^{i\alpha T_1} e^{i\beta T_2} e^{i\phi K_z}$ which comes out as
\begin{eqnarray}
L(u)=\left(
\begin{array}{cccc}
\cosh \phi +\left(\frac{\alpha^2}{2}+\frac{\beta^2}{2}\right)e^{-\phi} & 
\alpha & \beta & \sinh \phi - \left(\frac{\alpha^2}{2}+\frac{\beta^2}{2}
\right)e^{-\phi}\\
\alpha e^{-\phi} & 1 & 0 & -\alpha e^{-\phi}\\
\beta e^{-\phi} & 0 & 1 & -\beta e^{-\phi} \\
\sinh \phi +\left(\frac{\alpha^2}{2}+\frac{\beta^2}{2}\right)e^{-\phi} & 
\alpha & \beta & \cosh \phi - \left(\frac{\alpha^2}{2}+\frac{\beta^2}{2}
\right)e^{-\phi}
\end{array}
\right)\,,
\label{lu1}
\end{eqnarray}
where we have used the relation
$$e^{-\phi}=\cosh \phi - \sinh \phi\,.$$ As $L(u)u_0=u$ where $u_0$
and $u$ are as given in Eq.~(\ref{conv}), we get the following set of equations
\begin{eqnarray}
\gamma_u &=& \cosh \phi +\left(\frac{\alpha^2}{2}+
\frac{\beta^2}{2}\right)e^{-\phi}\,,
\label{ludc1}\\
\gamma_u u_x &=& -\alpha e^{-\phi}\,,
\label{ludc2}\\
\gamma_u u_y &=& -\beta e^{-\phi}\,,
\label{ludc3}\\
\gamma_u u_z &=& -\sinh \phi -\left(\frac{\alpha^2}{2}+
\frac{\beta^2}{2}\right)e^{-\phi}\,.
\label{ludc4}
\end{eqnarray}
Adding Eq.~(\ref{ludc4}) and Eq.~(\ref{ludc1}) we get
$$e^{-\phi}=\gamma_u(1+u_z)\,.$$ Taking logarithm of both sides we get
\begin{eqnarray}
\phi = -\ln [\gamma_u(1 + u_z)]\,.
\label{phif}
\end{eqnarray}
Using the expression of $e^{-\phi}$ in Eq.~(\ref{ludc2}) and Eq.~(\ref{ludc3})
we get
\begin{eqnarray}
\alpha = -\frac{u_x}{1+u_z}\,,\,\,\,\,
\beta = -\frac{u_y}{1+u_z}\,.
\label{ab}
\end{eqnarray}
Remembering that $L(u)$ is a $4\time 4$ matrix its individual matrix
elements can be written as $L_{n,m}$. As in SR, we want to specify all
the matrix elements $L_{n,m}$ in terms of the velocity components
$u_i$ where $i=x,\,y,\,z$. There remains two matrix elements of
$L(u)$, $L_{1,4}$ and $L_{4,4}$, in Eq.~(\ref{lu1}) which requires
some dressing up before they can be expressed in terms of the velocity
components. Adding Eq.~(\ref{ludc1}) to $L_{1,4}=\sinh \phi -
\left(\frac{\alpha^2}{2}+\frac{\beta^2}{2} \right)e^{-\phi}$ we get
$L_{1,4}+\gamma_u=e^\phi$. Some trivial manipulation then yields
\begin{eqnarray}
L_{1,4}=-\frac{\gamma_u(u^2+u_z)}{1+u_z}\,.
\label{14}
\end{eqnarray}
In a similar fashion subtracting Eq.~(\ref{ludc4}) from the expression 
of $L_{4,4}=\cosh \phi - \left(\frac{\alpha^2}{2}+\frac{\beta^2}{2}
\right)e^{-\phi}$ we get
\begin{eqnarray}
L_{4,4}=\frac{\gamma_u(1-u^2+u_z-u_z^2)}{1+u_z}\,.
\label{44}
\end{eqnarray}
Ultimately using Eq.~(\ref{ludc1}) to Eq.~(\ref{ludc4}) and Eq.~(\ref{14})
and Eq.~(\ref{44}) one can write $L(u)$ purely in terms of the velocity
components as written in Eq.~(\ref{mglu}).
\section{$HOM(2)$ inverse transformations}
\label{app2}
The $HOM(2)$ inverse transformation is 
$$L^{-1}(u) = e^{-i\phi K_z} e^{-i\beta T_2} e^{-i\alpha T_1}\,.$$ The
form of the matrices $e^{-i\phi K_z}$, $e^{-i\beta T_2}$ and $
e^{-i\alpha T_1}$ can be obtained from Eq.~(\ref{t1t2}) and
Eq.~(\ref{phikz}) by the following replacements: $\alpha \to -\alpha$,
$\beta \to -\beta$ and $\phi \to -\phi$. Multiplying $ e^{-i\phi
  K_z}$, $e^{-i\beta T_2}$ and $e^{-i\alpha T_1}$ in the specific
order, as required for the inverse transformation, we get
\begin{eqnarray}
L^{-1}(u)=\left(
\begin{array}{cccc}
\cosh \phi +\left(\frac{\alpha^2}{2}+\frac{\beta^2}{2}\right)e^{-\phi} & 
-\alpha e^{-\phi} & -\beta e^{-\phi} & -\sinh \phi - \left(\frac{\alpha^2}{2}
+\frac{\beta^2}{2} \right)e^{-\phi}\\
-\alpha e^{-\phi} & 1 & 0 & \alpha\\
-\beta & 0 & 1 & \beta \\
-\sinh \phi +\left(\frac{\alpha^2}{2}+\frac{\beta^2}{2}\right)e^{-\phi} & 
-\alpha e^{-\phi}& -\beta e^{-\phi} & \cosh \phi - 
\left(\frac{\alpha^2}{2}+\frac{\beta^2}{2}
\right)e^{-\phi}
\end{array}
\right).
\label{luinv1}
\end{eqnarray}
Again using Eq.~(\ref{ludc1}) to Eq.~(\ref{ludc4}) and Eq.~(\ref{14})
and Eq.~(\ref{44}) one can write $L^{-1}(u)$ purely in terms of the
velocity components as written in Eq.~(\ref{luinv}). It can easily be checked
that the resulting inverse $HOM(2)$ transformations satisfy
$$L^{-1}(u) u = u_0\,,$$ where $u_0$ and $u$ are as given in
Eq.~(\ref{conv}).

\end{document}